\documentclass[twocolumn]{aastex631}

\usepackage{savesym}
\usepackage{graphicx}	% Including figure files
\usepackage{amsmath}	% Advanced maths commands
\usepackage{amssymb}	% Extra maths symbols

\renewcommand{\figurename}{Fig.}

\savesymbol{longtable}
\usepackage{xcolor}

\usepackage[utf8]{inputenc}
\usepackage[english]{babel}

\restoresymbol{SUB}{longtable}

\usepackage{physics}

\begin{document}

\title{Suppression of the TeV pair-beam plasma instability by a tangled weak intergalactic magnetic field}

\author{Mahmoud Alawashra}
\affiliation{Institute for Physics and Astronomy, University of Potsdam, D-14476 Potsdam, Germany}
\email{mahmoud.s.a.alawashra@uni-potsdam.de}

\author[0000-0001-7861-1707]{Martin Pohl}
\affiliation{Institute for Physics and Astronomy, University of Potsdam, D-14476 Potsdam, Germany}
\affiliation{Deutsches Elektronen-Synchrotron DESY, Platanenallee 6, 15738 Zeuthen, Germany}
\email{martin.pohl@desy.de}

\received{04 Nov 2021}
\revised{04 Feb 2022}
\accepted{01 Mar 2022}
\published{-}
\submitjournal{AAS The Astrophysical Journal}

\begin{abstract}

We study the effect of a tangled sub-fG level intergalactic magnetic field (IGMF) on the electrostatic instability of a blazar-induced pair beam. Sufficiently strong IGMF may significantly deflect the TeV pair beams, which would reduce the flux of secondary cascade emission below the observational limits. A similar flux reduction may result from the electrostatic beam-plasma instability, which operates the best in the absence of IGMF. Considering IGMF with correlation lengths smaller than a  kpc, we find that weak magnetic fields increase the transverse momentum of the pair beam particles, which dramatically reduces the linear growth rate of the electrostatic instability and hence the energy-loss rate of the pair beam. We show that the beam-plasma instability is eliminated as an effective energy-loss agent at a field strength three orders of magnitude below that needed to suppress the secondary cascade emission by magnetic deflection. For intermediate-strength IGMF, we do not know a viable process to explain the observed absence of GeV-scale cascade emission.

\end{abstract}


\keywords{gamma rays: general -- instabilities -- magnetic fields  -- relativistic processes -- waves}

%%%%%%%%%%%%%%%%%%%%%%%%%%%%%%%%%%%%%%%%%%%%%%%%%%

%%%%%%%%%%%%%%%%% BODY OF PAPER %%%%%%%%%%%%%%%%%%

\section{Introduction}

GeV-TeV gamma-ray signals from various blazars ($z>0.024$) have been observed by the Fermi-LAT telescope and the imaging atmospheric Cerenkov telescopes (i.e VERITAS, MAGIC and HESS) \citep{1752,2010A&A...520A..83H}. Very high energy gamma rays annihilate with the extra-galactic background light (EBL), producing a collimated beam of electron-positron pairs, that are expected to quickly lose their energies via the inverse Compton scattering on the cosmic microwave background (CMB) \citep{1967PhRv..155.1408G,1970RvMP...42..237B}. Primary gamma rays of a few TeV would produce an electromagnetic cascade in the GeV energy band, but that emission appears to be absent in the gamma-ray spectra from some blazars \citep{2009PhRvD..80l3012N}.

One possible explanation for the absence of the GeV cascade emission is significant magnetic deflection of the electrons and the positrons of the beam \citep{PhysRevD.80.023010,PhysRevD.80.123012,2010Sci...328...73N,2011A&A...529A.144T,Takahashi_2011,Vovk_2012}. This deflection results in an extended emission or/and a time delay of the cascade emission. The field strength required to suppress the cascade emission due to the time delay is around $B_\mathrm{IGM}>10^{-16}\ \mathrm{G}$ for IGMF with a correlation length similar to or larger than the energy loss length of the beam, $\lambda_\text{B} \gtrsim 10$ kpc, and stronger than that for a small correlation length, for which the beam sees a fluctuating magnetic field and the deflection becomes diffusive \citep{Ackermann_2018}. Those fields might be the original form of seed fields that may be amplified to stronger magnetic fields in the galaxies and galaxy clusters \citep{Durrer_2013,Vachaspati:2020blt,Batista:2021rgm}.

If the magnetic field is strong enough to deflect by a radian or more, then the cascade emission from active galactic nuclei (AGN) with oblique jets (jets more than 30$^\circ$ off of our line of sight) should become visible \citep{Broderick:2016akd,2020ApJ...892..123T}, but corresponding emission has not been found \citep{Tiede:2017xng,Broderick_2018,Ackermann_2018}.

Another possibility are beam-plasma instabilities that work as an alternative energy loss mechanism to the inverse Compton scattering of the pair beam \citep{Broderick_2012,Schlickeiser_2012,Miniati_2013,Schlickeiser_2013,2014ApJ...790..137B,Sironi_2014,Chang_2014,Supsar_2014,Chang:2016gji,2016A&A...585A.132K,2017A&A...607A.112R,Vafin_2018,Vafin_2019,AlvesBatista:2019ipr,shalaby_broderick_chang_pfrommer_puchwein_lamberts_2020}. The beam-plasma instabilities involve both electrostatic and electromagnetic modes, the two stream-instability- ($\boldsymbol{k}\cross{\boldsymbol{E_{1}}}=0$ where $\boldsymbol{E_1}$ is the perturbed electric field) and transverse Weibel and filamentation modes ($\boldsymbol{k}\cdot{\boldsymbol{E_{1}}}=0$) \citep{doi:10.1063/1.3514586}. However, for the blazar-induced TeV pair beams the electrostatic modes dominate the wave spectrum \citep{2005PhRvE..72a6403B}, and Weibel-type modes are likely suppressed \citep{2017A&A...607A.112R}. Hence considering only the electrostatic oblique modes (wave vectors with finite angle to the beam propagation direction) recovers the essential physics \cite{Chang:2016gji}.

Through their nonlinear feedback these electrostatic waves transfer energy from the beam particles to heat in the intergalactic medium. Cosmological simulations including this heating process can successfully reproduce the observed IGM temperature and the effective optical depth as a function of redshift as well with several other observations \citep{Puchwein:2011sx}. \citet{Perry_2021} argued otherwise, namely that the back reaction of the unstable waves on the pair beam particles distribution moderately scatters the beam particles and does not impose a significant energy loss. We do not discuss the particulars of the nonlinear feedback here and leave this issue for future studies. 

The beam-plasma electrostatic instability operates best in the absence of a magnetic field. Noting that magnetic deflection needs more than a femto-Gauss field amplitude, here we address the effect on the electrostatic instability that would be imposed by much weaker intergalactic magnetic fields with a correlation length much smaller than the beam energy loss length. In particular, we investigate whether the plasma instability still is the dominant energy-loss process and how strongly the cascade emission is suppressed \citep{2019ApJ...870...17Y}.

In this work, we consider an IGMF with small correlation length far below the energy-loss length of the pair beam, $\lambda_\text{B}$ $<<$ $\lambda_e$, which deflects the electrons and positrons equivalently. Note that this condition implies that we assume the intergalactic magnetic fields to have no large-scale ($\gg$~kpc) or homogeneous component. We only consider the fluctuation component. Magnetic fields with strengths of $B_\mathrm{IGM} \ll 10^{-12}$ Gauss do not modify the linear dispersion relation of the beam-plasma instability obtained by the electrostatic approximation. However, those fields may impact the instability linear growth rate by their effect on the beam distribution function. The case of large magnetic-field correlation lengths involves a net current in the beam and will be the considered in a future publication.

Our IGMF model is the same as that widely used in the analysis of deflection and time-delay limits \citep{PhysRevD.80.023010,PhysRevD.80.123012,2010Sci...328...73N,2011A&A...529A.144T,Takahashi_2011,Vovk_2012}. The focus lies on a weaker field strength and on small correlation lengths. In such magnetic fields the electrons and the positrons of the blazar-induced pair beam perform a random walk passing through many regions with different field orientations, resulting in an increased angular spread of the pair beam that scales with the mean field strength and the square root of the correlation length \citep{Durrer_2013}.

We showed that this widening of the beam significantly slows the electrostatic instability, which decreases the energy loss rate of the beam particles. At a certain limit in the parameter space ($B_\mathrm{IGM},$ $\lambda_\text{B}$), driving the waves becomes less effective than inverse-Compton scattering the CMB, and the GeV cascade emission can no longer be suppressed. For the plasma instability model in \cite{Vafin_2018}, this limit is found to be around three orders of magnitude below the one that by magnetic deflection would impose a time delay of the cascade emission by 10 years \citep{Ackermann_2018}. 

The structure of this paper is as follows. In section \ref{sec:2}, we present the linear growth rate spectrum of the electrostatic instability of a realistic pair beam distributions without and with weak intergalactic magnetic fields. In section \ref{sec:3}, we present the nonlinear instability saturation of the unstable electrostatic waves. Finally, we demonstrate our results in section \ref{sec:result} and conclude in section \ref{sec:con}.

\section{Linear Growth Rate of the Electrostatic Instability}\label{sec:2}

In this section, we present the linear growth rate of electrostatic waves for a realistic blazar-induced pair beam with finite angular spread (kinetic instability) moving in an unmagnetized intergalactic medium. Then we consider the magnetic fields in the intergalactic medium and find their impact on the beam distribution function and the implications for the growth rate of electrostatic waves.

As we mentioned in the introduction, the electrostatic approximation is valid for the blazar-induced pair beam for which the electrostatic modes grow far more quickly than do the electromagnetic modes \citep{2010PhRvE..81c6402B,Chang:2016gji}. A comparison of the Weibel growth rate for blazar-induced pair beams using a cold-beam distribution \cite{Schlickeiser_2012} and a Waterbag distribution \citep{2017A&A...607A.112R} shows that the Weibel instability is suppressed for a realistic blazar-induced pair beam. Therefore, we will proceed with the electrostatic approximation in our analysis.

Linearizing the Vlasov–Maxwell equations for electrostatic waves leads to the following dispersion relation \citep{1990}
\begin{equation}\label{eq:dis}
    1 - \frac{\omega_p^2}{\omega^2} - \sum_b \frac{4\pi n_b e_b^2}{k^2} \int d^3\boldsymbol{p} \frac{\boldsymbol{k}\cdot \frac{\partial f_b(\boldsymbol{p})}{\partial \boldsymbol{p}}}{\boldsymbol{k}\cdot\boldsymbol{v} - \omega}=0, 
\end{equation}

where $f_b(\boldsymbol{p}) = f_b(\boldsymbol{p},\boldsymbol{x})/n_b$ is the normalized momentum distribution function of the beam, $n_b$ is the total number density of the beam, $\omega_p = (4\pi n_e e^2/m_e)^{1/2}$ is the plasma frequency of the intergalactic background plasma with density $n_e$. The wave vector is chosen as $\boldsymbol{k} = (k_\perp,0,k_{||})$, and the beam propagates along the $z$ axis with cylindrical symmetry.

In our analysis we consider the kinetic instability for which the beam temperature plays a significant role. The kinetic instability is applicable, if the velocity spread times the wave vector of the unstable waves is larger than the growth rate of the reactive instability \citep{Chang:2016gji}
\begin{equation}\label{eq:react}
    \abs{\boldsymbol{k}\cdot \boldsymbol{\Delta v}} >> \omega_{i,\text{r}} .
\end{equation}
The peak reactive growth rate %along the longitudinal component of the wavevector to the beam direction using the electrostatic approximation 
is \citep{doi:10.1063/1.3514586} 
\begin{equation}
    \omega_{i,\text{r}} = \frac{\sqrt{3}}{2^{4/3}} \omega_p \left(\frac{n_b}{\gamma_b n_e}\right)^{1/3}\left(\left(\frac{k_\perp}{k}\right)^2+\frac{1}{\gamma_b^2}\left(\frac{k_{||}}{k}\right)^2\right)^{1/3},
\end{equation}
where $\gamma_b$ is the beam Lorentz factor, and the parallel wave number is fixed at the resonance, $k_{||}= \omega_p/c$. For a relativistic beam the perpendicular velocity spread is $\Delta v_\perp \approx \frac{c}{\gamma_b}$, and the parallel velocity spread is $\Delta v_{||} \gtrsim \frac{c}{\gamma_b^2}$, resulting from the Lorentz boost of the beam from the COM to the lab frame \citep{Miniati_2013}. For realistic blazar-induced pair beam with Lorentz factor $\gamma_b \sim 10^5 - 10^6$, condition (eq(\ref{eq:react})) is satisfied for essentially all oblique waves, meaning that we should consider the kinetic regime and not the reactive one (cold-limit).

For a relativistic electron beam ($\gamma_b >> 1$) with a small angular spread ($\Delta \theta << 1$ rad) traveling in a homogeneous background plasma with a number density $n_e$, the dispersion relation, eq.~\ref{eq:dis}, in the kinetic regime yields the following linear growth rate of electrostatic waves \citep{1990}

\begin{equation}\label{eq:growth}
\begin{split}
    \omega_i(k) = & \pi \omega_p \frac{n_b}{n_e} \left(\frac{\omega_p}{kc}\right)^3 \int_{\theta_1}^{\theta_2}d\theta \ \frac{\partial g(\theta)}{\partial \theta}\\ & \times  \frac{-2g(\theta)\sin{\theta+(\cos{\theta}-\frac{kc}{\omega_p}\cos{\theta'}})}{[(\cos{\theta_1}-\cos{\theta})(\cos{\theta}-\cos{\theta_2})]^{1/2}},
\end{split}
\end{equation}
where
\begin{equation}
    g(\theta) = m_e c \int_0^\infty dp \text{ } p f_b(p,\theta),
\end{equation}
and 
\begin{equation}
    \cos{\theta_{1,2}} = \frac{\omega_p}{kc} \left( \cos{\theta'}\pm \sin{\theta'} \sqrt{ \left(\frac{kc}{\omega_p}\right)^2-1}\right),
\end{equation}
where $k=\sqrt{k_\perp^2+k_{||}^2}$ is the module of the unstable electrostatic waves wave-number vector ($k_\perp$ and $k_{||}$ are the perpendicular and the parallel components to the beam propagation direction respectively), $\theta'$ is the angle between the wave vector and the beam propagation direction, and $\theta$ is the angle between the particle momentum and the beam direction axis ($z$-axis). The beam is azimuthally symmetric around the propagation axis. 
 
%\rev{, $\omega_p=(4\pi n_e e^2/m_e)^{1/2}$ is the plasma frequency of the background electrons, $m_e$ is the electron mass, and $c$ is the speed of light.} DELETED

The momentum distribution function, $f_b(p,\theta)$, of the beam is normalized as follows
\begin{equation}\label{eq:norm}
    2\pi\int_0^\infty dp \text{ } p^2 \int_0^{\pi} d\theta \text{ } \sin{\theta} f_b(p,\theta) = 1,
\end{equation}
and can be factorized into parallel and perpendicular components
\begin{equation}\label{eq:f_dis}
    f_b(p,\theta) = f_{b,p}(p)f_{b,\theta}(p,\theta),
\end{equation}
where for the parallel momentum distribution $f_{b,p}(p)$ we used eq.~26 and eq.~56 in \citet{Vafin_2018} that are obtained for a realistic pair beam at a distance of 50~Mpc from the blazar. The angular distribution, $f_{b,\theta}(p,\theta)$, depends on whether or not we have intergalactic magnetic fields. 

\subsection{Electrostatic instability for a pair beam in non-magnetized intergalactic medium}\label{sec:2.1}

\begin{figure}
\centering
    \includegraphics[width=8.0cm]{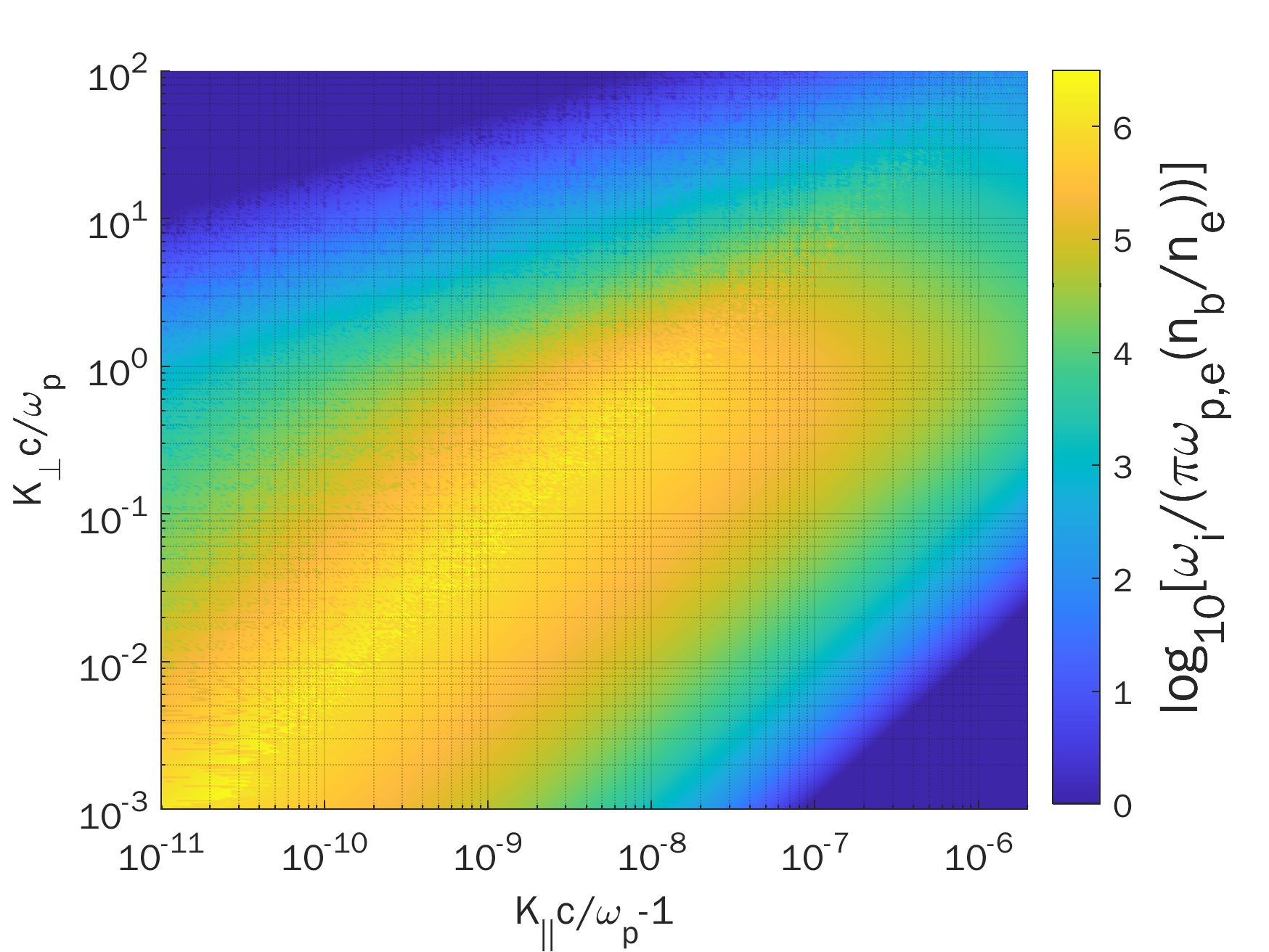}
    \caption{Normalized electrostatic growth rate in the absence of an external magnetic field.}
    \label{fig:wiB=0}
\end{figure}

In the case of a non-magnetized intergalactic medium, the angular spread of the beam is due to the angular energy spread only. In this case, the angular distribution function of the beam, $f_{b,\theta}(p,\theta)$, can be approximated by a Gaussian \citep{Miniati_2013}
\begin{equation}\label{eq:spread}
    f_{b,\theta}(p,\theta) \approx \frac{1}{\pi \Delta\theta_s^2} \exp{-\frac{\theta^2}{\Delta \theta_s^2}},
\end{equation}
where the angular energy spread approximated as \citep{Broderick_2012}
\begin{equation}\label{eq:ths}
    \Delta\theta_s \approx \frac{m_e c }{p }
\end{equation}
Substituting eq(\ref{eq:spread}) into eq(\ref{eq:f_dis}) and eq(\ref{eq:growth}) we found the numerical solution for the linear electrostatic growth rate as shown in Fig.\ref{fig:wiB=0}, we see that most of the unstable modes are in the oblique and the parallel directions. The maximum linear growth rate is found to be
\begin{equation}
\begin{split}
    \omega_{i,\text{max}}  &=   (3.83\times 10^{-7}) \omega_p \frac{n_{b20}}{n_{e7}},\\
    & = (1.15\times 10^{-8}) \omega_p, 
\end{split}
\end{equation}
where $\omega_p = 17.8$ Hz is the plasma frequency of the intergalactic background electrons for the unit density $n_{e}= n_{e7} 10^{-7}$cm$^{-3}$ = $10^{-7}$cm$^{-3}$. For the fiducial pair beam parameters, the number density of the pair beam is $n_{b}= n_{b20} 10^{-20}$cm$^{-3}$ = $3\times10^{-22}$cm$^{-3}$. Note that the maximum growth rate we found here is twice that reported in \citet{Vafin_2018}, because the maximum growth rate in \cite{Vafin_2018} was computed for the parallel wave numbers down to $(k_\parallel c /\omega_p - 1) \approx 10^{-7}$, however we found that smaller parallel wave numbers down to $(k_\parallel c /\omega_p - 1) \approx 10^{-14}$ have a larger growth rates, as shown in Fig.\ref{fig:wiB=0}.

\citet{Vafin_2018} demonstrated that for a blazar with a redshift $z=0.2$ those unstable waves drain the pair beam energy around a hundred times faster than does inverse-Compton scattering on the CMB, taking into account the modulation instability as a damping process. The main uncertainties in that work are the assumptions on the spectrum and gamma-ray flux from the blazar and the approximation of the nonlinear saturation level.

The growth rate is calculated for a pure electron beam moving in a background plasma of electrons and ions. \citet{Schlickeiser_2012} demonstrated that having separate distribution functions for electrons and for positrons yields the same growth rate as do calculations that assume only an electron beam \citep{Broderick_2012}.

\begin{figure*}
       \centering
        \begin{minipage}[b]{0.475\textwidth}
            \centering
            \renewcommand{\thefigure}{2a}
            \includegraphics[width=\textwidth]{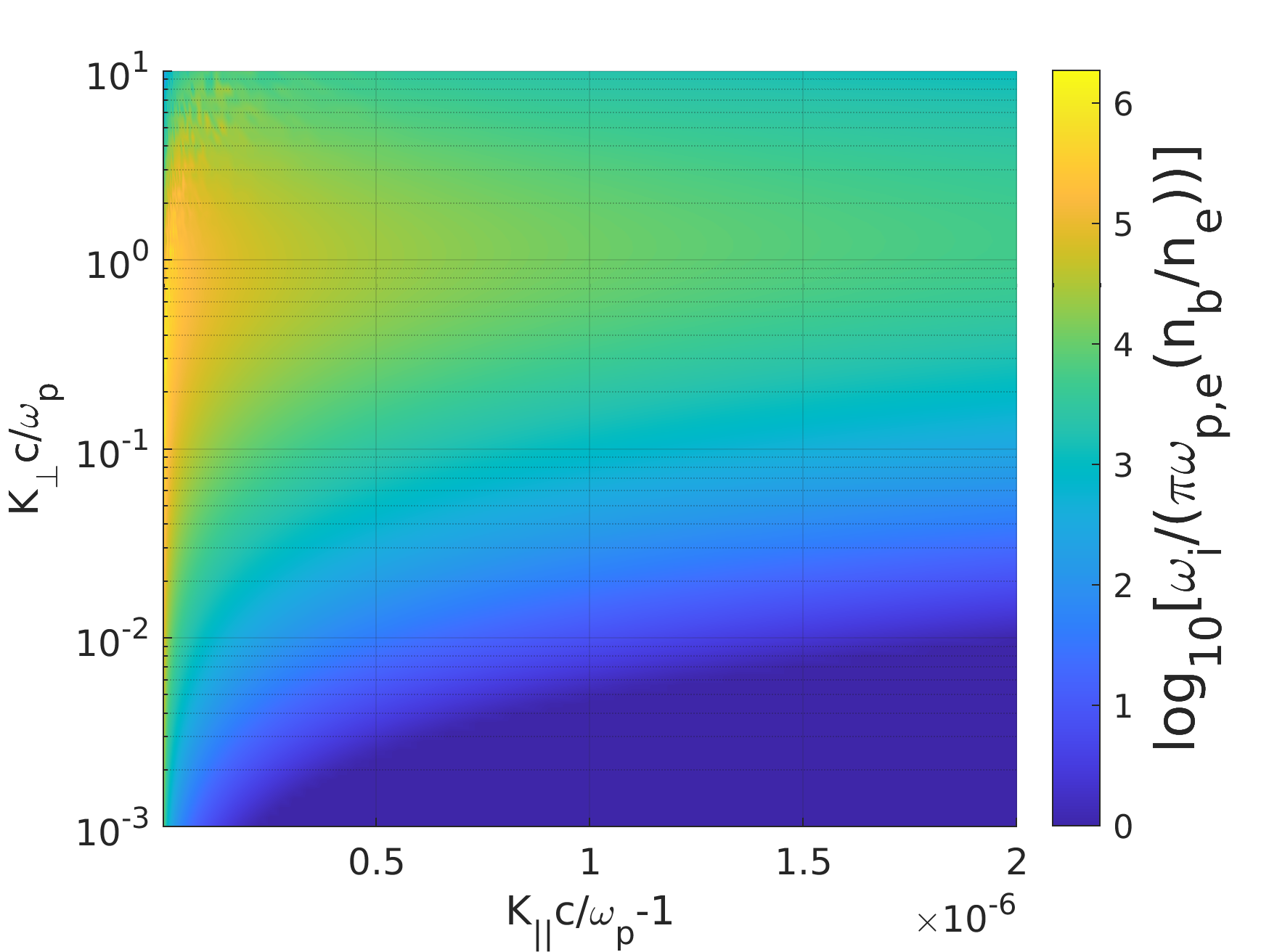}
            \caption{$B_{\mathrm{IGM}} = 10^{-18}$ Gauss and $\lambda_B = 1$ pc  }    
            \label{fig:b-18}
        \end{minipage}
        \hfill
        \begin{minipage}[b]{0.475\textwidth}  
            \centering 
            \renewcommand{\thefigure}{2b}
            \includegraphics[width=\textwidth]{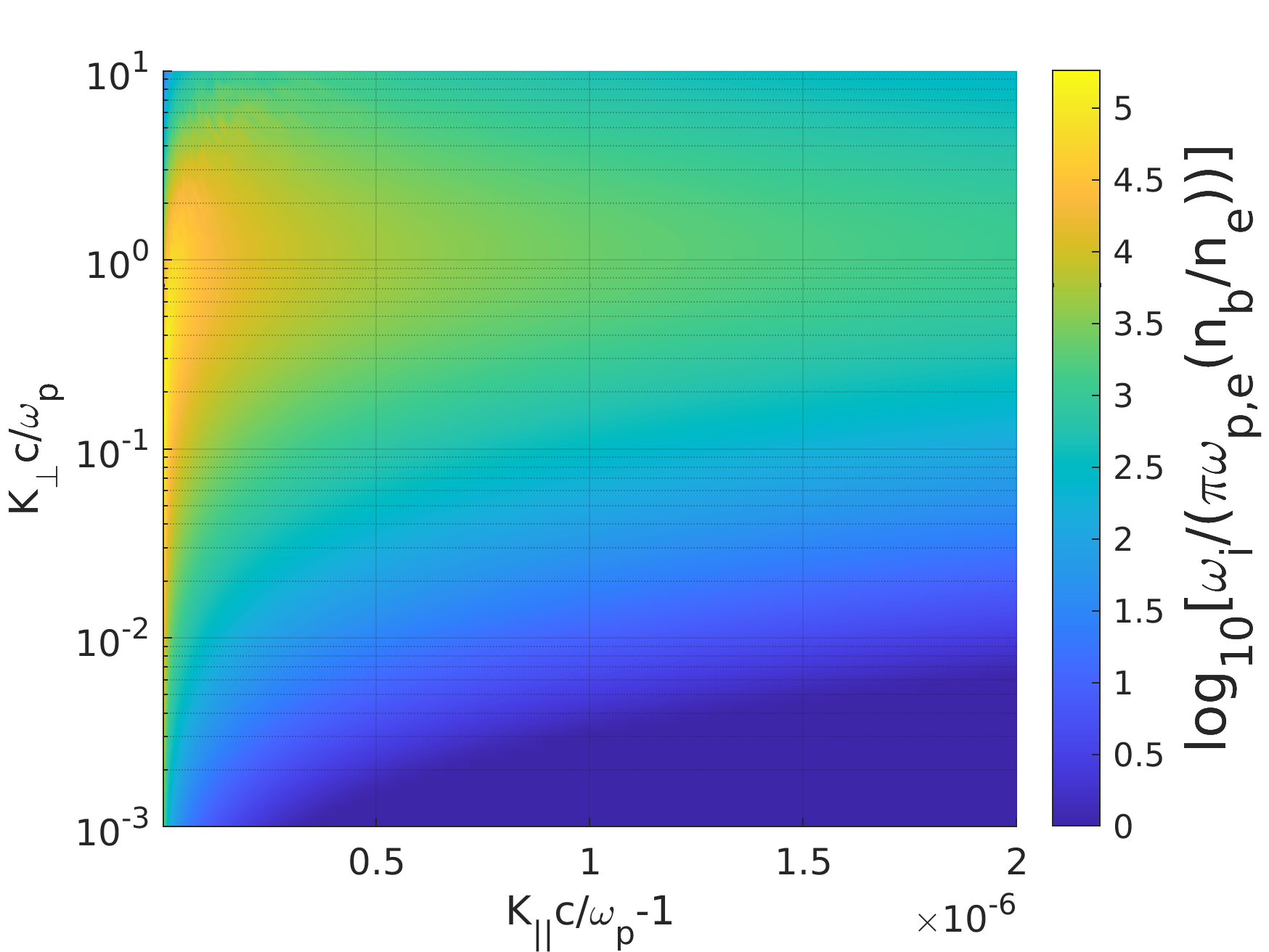}
            \caption{$B_{\mathrm{IGM}} = 10^{-17}$ Gauss and $\lambda_B = 1$ pc   }    
            \label{fig:b-17}
        \end{minipage}
        \vspace{5pt}
        \begin{minipage}[b]{0.475\textwidth}   
            \centering 
            \renewcommand{\thefigure}{2c}
            \includegraphics[width=\textwidth]{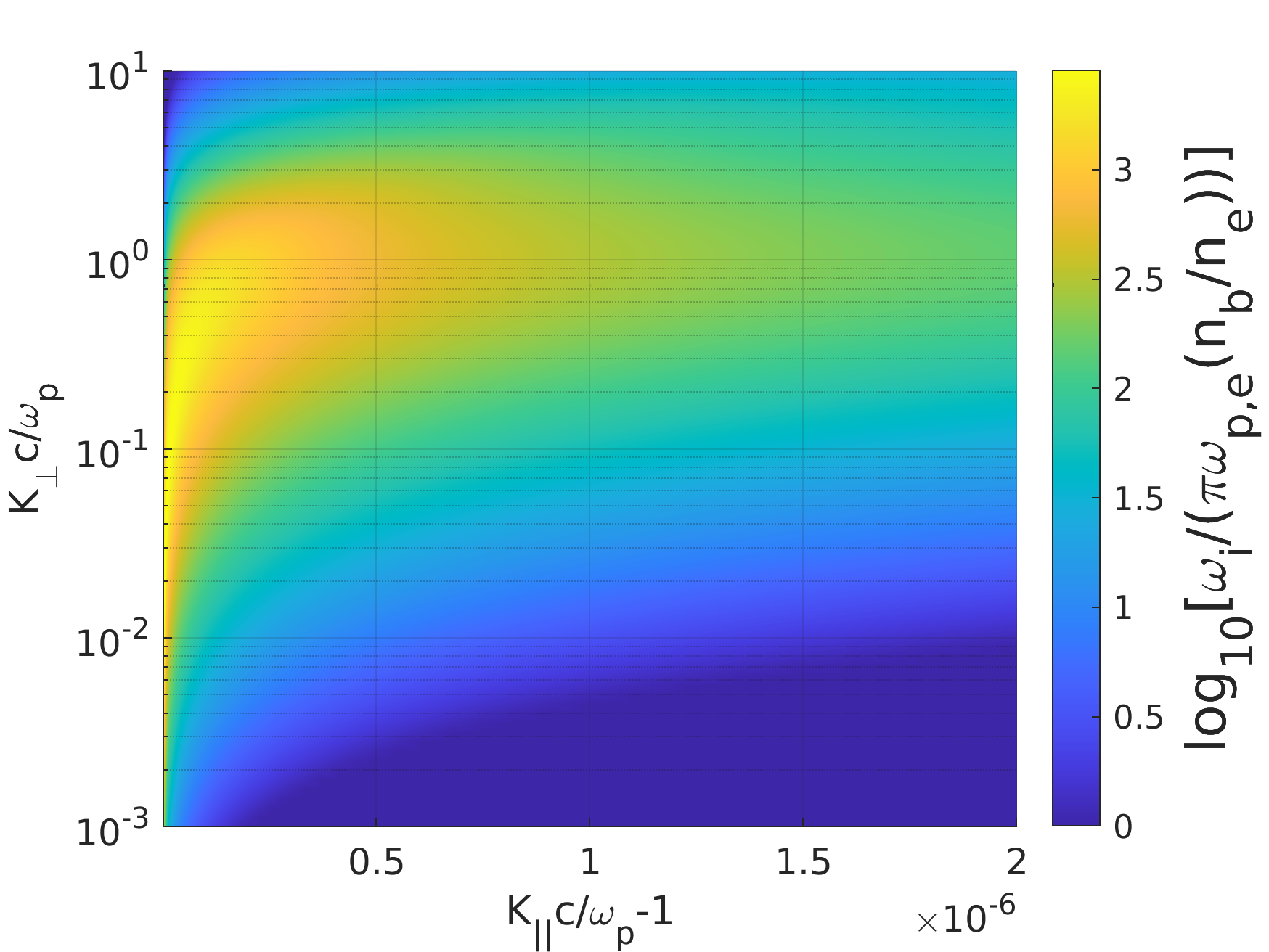}
            \caption{$B_{\mathrm{IGM}} = 10^{-16}$ Gauss and $\lambda_B = 1$ pc. }    
            \label{fig:b-16}
        \end{minipage}
        \hfill
        \begin{minipage}[b]{0.475\textwidth}   
            \centering 
            \renewcommand{\thefigure}{2d}
            \includegraphics[width=\textwidth]{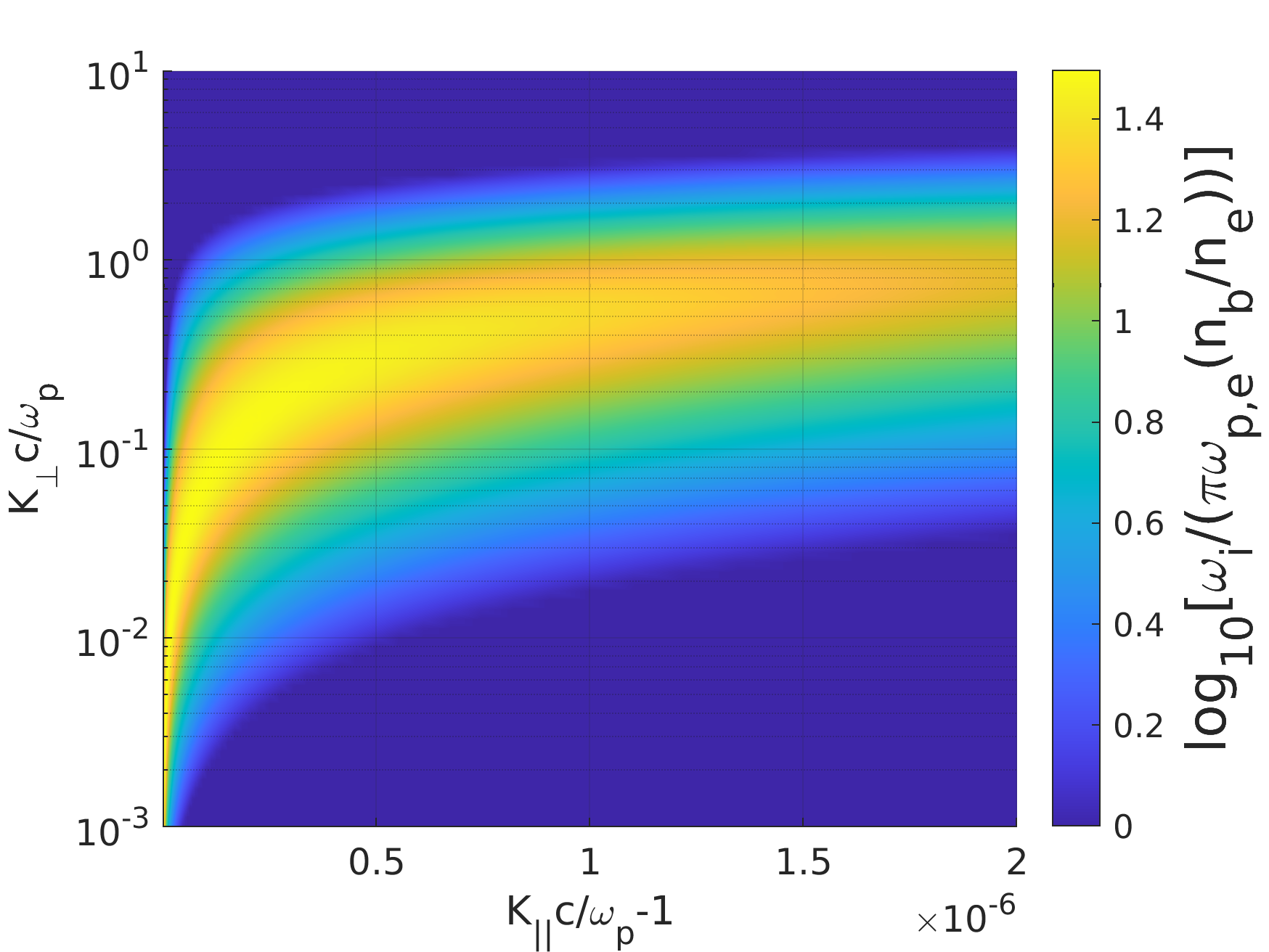}
            \caption{$B_{\mathrm{IGM}} = 10^{-15}$ Gauss and $\lambda_B = 1$ pc.} 
            \label{fig:b-15}
        \end{minipage}
        \renewcommand{\thefigure}{2}
        \caption{The longtime of the normalized electrostatic growth rate ($\log_{10}(\omega_i /(\pi\omega_{p,e}(n_b/n_e)))$) for different intergalactic magnetic field strength values, we see that as the intergalactic magnetic fields strength increases the linear growth rate of the instability growth rate decreases.} 
        \label{fig:Bwi}
    \end{figure*}

\subsection{Electrostatic instability for a pair beam with a weak intergalactic magnetic field} \label{sec:2.2}

We address in this section the effects of weak intergalactic magnetic fields on the electrostatic plasma instability. If the electron gyromagnetic frequency, $\omega_B = eB_\mathrm{IGM}/m_e$, is much smaller than their plasma frequency, $\omega_B \ll \omega_p$, then an external magnetic field doesn't change the electrostatic dispersion relation used to derive the linear growth rate \citep{osti_4815657}. The corresponding upper limit for the strength of the intergalactic magnetic field is $B_\mathrm{IGM}\lesssim 10^{-9}$ Gauss, where we again assumed the number density to be $n_e = 10^{-7}\,$cm$^{-3}$.

The magnetic-field correlation lengths we consider, $\lambda_\text{B}\sim 10^{3} - 10^{-5}$~pc, are much larger than the intergalactic plasma skin length, $\lambda_D \sim 5\times 10^{-10}$~pc, meaning that even the variations of the IGMF have no direct impact on the beam plasma dispersion relation. However, the directional changes of the magnetic field affect the equilibrium beam distribution function, which in turn impacts on the linear electrostatic growth rate eq(\ref{eq:growth}). In other words, the blazar-induced pair beam that triggers the instability travels through many correlations lengths in the IGMF. For example, the blazar-induced pair beam distribution function we are considering is this work is calculated at distance 50 Mpc in the IGM from the blazar \citep{Vafin_2018}, whereas the pair production starts at distances smaller than 1 Mpc \citep{Miniati_2013}.

More importantly, we can take the inverse Compton scattering length, $\lambda_\text{IC} \approx 75\,\text{kpc} \left({10^7}/{\gamma_b}\right)$, as an upper limit on the energy loss length of the beam particles, which gives around 188 kpc for a Lorentz factor of $\gamma_b = 4\times 10^{6}$. This means that the pair beam distribution function carries the effects of the magnetic fields over a large number of directional changes, since most of the particles in the beam have traveled many correlation lengths at least, $\lambda_\text{IC}>>\lambda_\text{B}$. This propagation of the pair beam over many correlation length imposes an additional angular spread on its momentum distribution which in turn significantly affects the linear electrostatic growth rate.

%We explicitly consider an intergalactic magnetic field with a correlation length that is much smaller than the energy-loss length of the pair beam due to inverse-Compton scattering, $\lambda_B<<\lambda_\mathrm{IC}$. 

Those fields lead to stochastic deflections of the electrons and positrons that diffusively widen the angular distribution function of the pair beam as shown in appendix \ref{app:G}.  Adding in quadrature the energy angular spread $\Delta\theta_s$ (eq.~\ref{eq:ths}) and the magnetic widening $\Delta\theta_\text{IGMF}$ (eq.~\ref{eq:thIMGF}) gives the following distribution of the angular spread of the pair beam after travelling many correlation lengths in the IGM
\begin{equation}\label{eq:fth_dis}
    f_{b,\theta}(\theta,p)=\frac{1}{\pi \Delta\theta^2}\exp{-\big(\frac{\theta}{\Delta\theta}\big)^2}, \quad \text{             } 0\leq\theta\leq\pi,
\end{equation}
where 
\begin{equation}\label{eq:a}
    \Delta\theta = \frac{m_e c}{p} \sqrt{1 + \frac{2}{3}\lambda_B\lambda_{\text{IC}}\left(\frac{e B_{\mathrm{IGM}}}{m_e c}\right)^2}.
\end{equation}

Note that the result in appendix \ref{app:G} for the magnetic deflection, $\Delta\theta_\text{IGMF}$, is consistent with the diffusion angle used in the IGMF deflection analyses, e.g. eq.~31 in \citet{2009PhRvD..80l3012N}. 

\begin{figure}
    \centering
    \renewcommand{\thefigure}{3}    
    \includegraphics[width=3.5in]{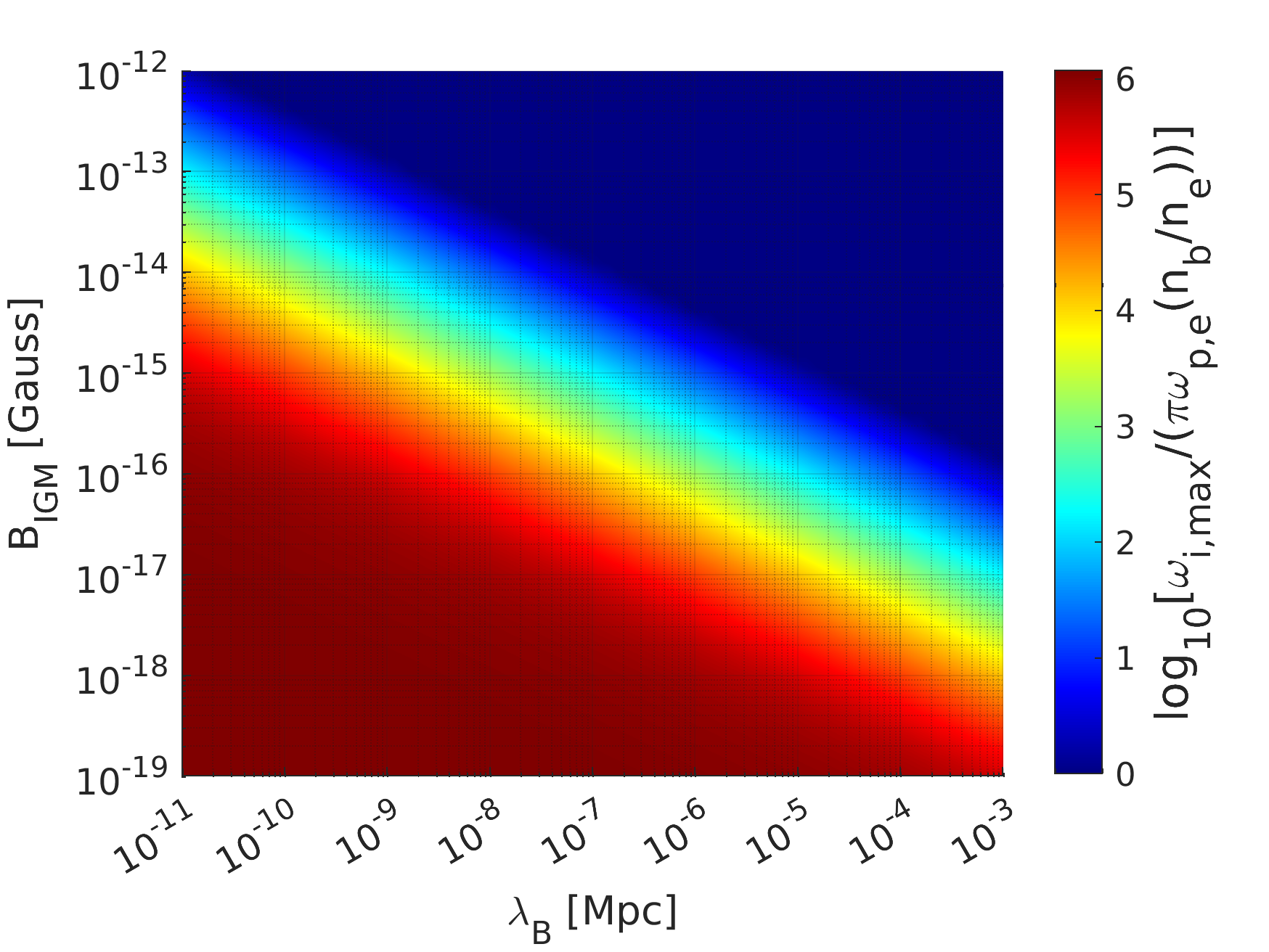}
    \caption{The logarithm of the maximum normalized growth rate, $\log_{10}(\omega_{i,\text{max}} /(\pi\omega_{p,e}(n_b/n_e)))$.}
    \label{fig:maxwi}
\end{figure}

Finally, substituting eq(\ref{eq:fth_dis}) into eq(\ref{eq:f_dis}) and eq(\ref{eq:growth}) we numerically found the linear growth rate spectrum for a few values of the IGMF strength, $B_{\mathrm{IGM}}$, and the correlation length, $\lambda_B$, and displayed it in Fig.\ref{fig:Bwi}. The main impact of the IGMF is a general reduction of the growth rate. Fig.\ref{fig:maxwi} shows the peak growth rate as a function of $B_{\mathrm{IGM}}$ and $\lambda_B$. To be noted from the figure is that specific values of the peak growth rate are found on a characteristic $B_{\mathrm{IGM}}\propto \lambda_B^{-0.5}$. The reduction of the instability growth rate increases the energy loss time due to the plasma instability as we will see in the next section. 

%Notice also that the magnetic field should be constant over many plasma wavelengths for the dispersion relation to be able to describe the system. Since the plasma wavelength of the intergalactic background electrons is around $10^{-9}$ pc then it's safe to consider magnetic fields with a spacial scales up to $10^{-5}$ pc. 

\section{Nonlinear instability saturation} \label{sec:3}

The unstable electrostatic waves grow exponentially with the linear growth rate, accumulating at the resonant parallel wave number $k_{||} \approx \omega_p/c$. Depending on their nonlinear interactions, the waves could drain the kinetic energy of the pair beam and heat the IGM. The first type of those nonlinear interactions is the scattering of the electrostatic waves on the background plasma, known as nonlinear Landau interactions. The second nonlinear interaction is a wave-wave interaction between the electrostatic waves known as the modulation instability. The first process operates at any wave intensity, whereas the second occurs only above a certain threshold. 

Simulations of the evolution of the beam/plasma system are impossible right now for realistic parameters. However, there are various analytical estimates in the literature concerning the energy density that the waves reach in an equilibrium state \citep{Miniati_2013,Schlickeiser_2012,Broderick_2012,Vafin_2018}. The inverse energy loss time of the pair beam due to the electrostatic instability is given by \citep{Vafin_2018,Miniati_2013}
\begin{equation}\label{eq:loss}
    \tau_\text{loss}^{-1} = 2 \delta \omega_{i,\text{max}},
\end{equation}
where $\omega_{i,\text{max}}$ is the peak linear growth rate and $\delta=U_\text{ES}/U_\text{beam}$ is the normalized wave energy density at the equilibrium level. The reduction of the linear growth rate due to the IGMF translates into an increase of the energy loss time. At some limit the beam-plasma instability becomes less relevant than the inverse Compton scattering. We will find this limit in the next section.

The wave intensity, $\delta$, depends on the non-linear evolution of the electrostatic waves, for which have different estimates. In the next section, we are going to include first the result given in \cite{Vafin_2018} and then discuss the implications of changing the value of the intensity of the waves to that found by \cite{Broderick_2012}.

\section{Results}\label{sec:result}

We found the maximum linear growth rate of the unstable electrostatic 2D spectrum for each intergalactic magnetic field strength, $B_\mathrm{IGM}$, and correlation length, $\lambda_B$, as shown in Fig.\ref{fig:maxwi}. Then we calculated the approximated energy loss time of the beam based on the maximum linear growth rate as in eq(\ref{eq:loss}), using the intensity of the waves given in \cite{Vafin_2018}, $\delta = 10^{-5}$. This time should be smaller than the inverse Compton scattering energy loss time, otherwise the beam-plasma instability cannot suppress the secondary cascade. The energy loss time of the inverse Compton scattering is given by
\begin{equation}\label{eq:IC}
    \tau_\text{IC} = \left(7.7\times 10^{13}\ \mathrm{s}\right) \, (1+z)^{-4} \left(\frac{10^6}{\gamma_b}\right),
\end{equation}
which at redshift $z=0.2$ and for a pair-beam Lorentz factor $\gamma_b = 4 \times 10^6$ gives the following ratio for the beam-plasma instability loss time given in \cite{Vafin_2018}
\begin{equation}\label{eq:comp}
    \frac{\tau_\text{loss}}{\tau_\text{IC}} = 0.026,
\end{equation}
if the intergalactic magnetic field is zero.

\begin{figure}
    \centering
    \renewcommand{\thefigure}{4}
    \includegraphics[width=8.5cm]{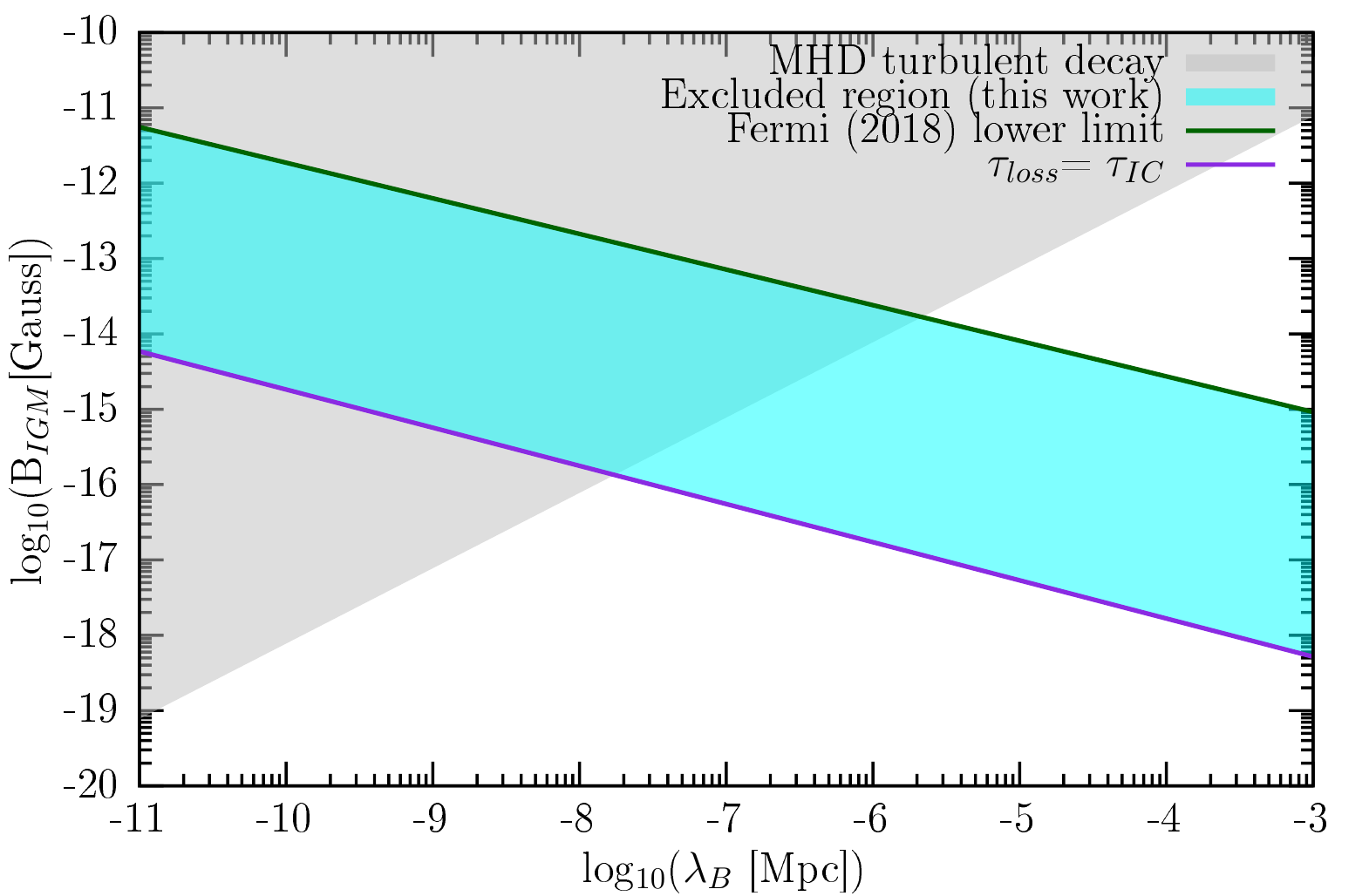}
    \caption{The excluded region of the IGMF, for which neither magnetic deflection nor the oblique instability can explain the absence of cascade emission. The gray region is the upper limit on the intergalactic magnetic fields strength due to the MHD turbulent decay \citep{PhysRevD.70.123003,Durrer_2013}.
    }
    \label{fig:Result}
\end{figure}

Using eq.~\ref{eq:loss}, we infer that a reduction by a factor 40 of the instability growth rate is sufficient to render it inefficient. The dependence of the growth rate on $B_\mathrm{IGM}$ and $\lambda_B$ (cf.  Fig.~\ref{fig:maxwi}) can then be turned into a limit in the $B_\mathrm{IGM}$-$\lambda_B$ parameter space, above which inverse Compton scattering provides the dominant energy loss of the pair beam. We show this limit in Fig.\ref{fig:Result}. It is at the same time an exclusion limit, because the then unavoidable inverse-Compton emission is not seen, and so the cyan-shaded are in the figure is excluded for the IGMF. For weaker field, the oblique instability may drain the beam energy sufficiently quickly, and for stronger fields the time delay of the cascade emission causes substantial uncertainty in the interpretation of the Fermi-LAT data of GeV-scale cascade emission. 

We see in Fig.\ref{fig:Result} that the beam-plasma instability suppression limit (the purple line) is three orders of magnitude lower than the lower limit on the IGMF strength needed to impose a significant time delay of the cascade emission flux due to the magnetic deflection (the green line) \citep{Finke_2015,Ackermann_2018,2011A&A...529A.144T}. We follow \cite{Ackermann_2018} in assuming an time period of 10 years as sufficient for the suppression of the cascade signal. The actual deflection angle would be well below one minute of arc.

Finally, to account for the uncertainty of the non-linear saturation level of the waves, we consider also the beam-plasma instability model presented in \cite{Broderick_2012} with $\delta =0.2$ and check how the plasma instability suppression limit changes for a certain correlation length. For $\lambda_\text{B} =10^{-11}$ Mpc, the instability limit would shift for \cite{Broderick_2012} to $B_\mathrm{IGM}=10^{-12.5}$ Gauss which is two orders of magnitude higher than that based on \cite{Vafin_2018} but still below the Fermi time delay lower limit. This shift applies also to all the instability suppression limit points in Fig.\ref{fig:Result}, since the correlation length and the intergalactic magnetic field strength determine together the angular spread eq.\ref{eq:a} that plays the key role in determining the linear growth rate. 

Although the nonlinear saturation level had changed by four orders of magnitudes between the models of \citet{Vafin_2018} and \citet{Broderick_2012}, the magnetic field limit had changed by only two orders of magnitude. This is due to the dependence of the energy loss time on the angular spread as $\tau_\text{loss} \propto (\Delta \theta)^{2}$, which is a result of the maximum linear growth rate dependence on the angular spread as $\omega_{i,\text{max}} \propto (\Delta \theta)^{-2}$ \cite{Vafin_2019}, and the energy loss time relation with the linear growth rate as $\tau_\text{loss} \propto \omega_{i,\text{max}}^{-1}$ eq(\ref{eq:loss}).

For a generic beam-plasma instability model with plasma instability energy loss time $\tau_\text{loss,0}$ in the absence of the IGMF, the energy loss time increases as $\tau_\text{loss} \propto (\Delta \theta)^{2}$ when the angular spread increases with the IGMF strength and correlation length as in eq.\ref{eq:a} reaching the inverse Compton scattering energy loss time at the following intergalactic magnetic field strength
\begin{equation}\label{eq:limitB}
\begin{split}
    \log{\left(\frac{B_\text{IGM,lim}}{\text{Gauss}}\right)} = &-17.92 -\frac{1}{2}\log{\left(\frac{\lambda_\text{B}}{\text{pc}}\right)}  \\ & + \log{\left(\sqrt{\frac{\mathrm{Myr}}{\tau_{\text{loss},0}}}-\sqrt{\frac{\mathrm{Myr}}{\tau_\text{IC}}}\right)}.
\end{split}
\end{equation}
Eq(\ref{eq:limitB}) provides the intergalactic magnetic fields strength that is sufficient to suppress a general plasma instability, with energy loss time $\tau_\text{loss,0}$ in the absence of the IGMF, against the inverse Compton scattering of the blazar-induced pair beam on the CMB. For Vafin's model, the last logarithmic term on the right-hand side of eq.~\ref{eq:limitB} has a value very close to unity.

\section{Conclusion}\label{sec:con}

We investigated the effects of tangled weak intergalactic magnetic fields with small correlation lengths on the electrostatic instability driven by blazar-induced pair beams. The weak fields increase the angular spread of the pair beam which decreases the linear growth rate of the electrostatic beam-plasma instability, which in turn reduces the associated energy loss rate. 

In a certain region in the $B_\mathrm{IGM}$-$\lambda_B$ parameter space, neither the beam-plasma instability nor the intergalactic magnetic field deflection can explain the absence of cascade emission in the spectra of some TeV blazars, and so this parameter space region can be excluded, unless there is a third mechanism that suppresses the GeV-band cascade. 

Considering the beam-plasma instability model of \cite{Vafin_2018}, we can exclude an IGMF strength within the three orders of magnitude below the limit above which magnetic deflection imposes a significant time delay of the cascade (ten years). Even for the non-linear evolution model of \cite{Broderick_2012}, we can exclude a range a values that is one order of magnitude wide.

%This result indicates that the electrostatic plasma instability does not affect the lower limits on the magnetic fields needed to suppress the cascade emission by magnetic deflection. Also, our result exclude the parameter region between the instability suppression limit and the time-delayed lower bound in Fig.\ref{fig:Result} since in this parameter space region neither the plasma instability nor the intergalactic magnetic fields deflection work as an efficient suppressing mechanism of the GeV secondary cascade. 

Although the parameter space below the beam-plasma instability suppression limit is not excluded by the cascade observations, part of this region (at $\lambda_\text{B}\lesssim 10^{-8}\ \mathrm{Mpc}$ and shaded in gray in Fig.\ref{fig:Result}) is constrained by MHD turbulent decay \citep{PhysRevD.70.123003,Durrer_2013}. In conclusion, the allowed region for the IGMF lies below $10^{-16}$ Gauss at $\lambda_\text{B}\approx 10^{-8}$~Mpc, and the constraint on the IGMF strength is tighter than that for both larger correlation lengths (due to the instability suppression) and smaller correlation lengths (due to the MHD turbulent decay).

\section*{Acknowledgement}
This work was supported by the International Helmholtz-Weizmann Research School for Multimessenger Astronomy, largely funded through the Initiative and Networking Fund of the
Helmholtz Association. We thank and acknowledge Andrew Taylor for his helpful comments on our paper. 

\text{

}

%%%%%%%%%%%%%%%%% APPENDICES %%%%%%%%%%%%%%%%%%%%%

\appendix

\section{Beam distribution function with IGMF}\label{app:G}

Consider a magnetic field with a constant magnitude that arbitrarily and abruptly changes its direction every correlation length, $\lambda_{B}$, along the beam propagation line. In Cartesian coordinates with the $z$ axis aligned with the beam, the magnetic-field component in the $x-y$ plane deflects the beam every $\lambda_{B}$ interval in a different direction. We will include first the deflection due to the magnetic field component along the $x$ axis then we will generalize to the $x-y$ plane. At the end we find that the angular distribution function is a Gaussian with azimuthal symmetry, hence the electrons and positrons distribution functions are equivalent, and it is sufficient to consider only one species. 

At a given correlation-length interval denoted by $i$, the component of the magnetic field in the $x$ direction ($B_{x,i} = B_{\mathrm{IGM}} \sin{\theta'}\cos{\varphi'}$) deflects the beam positrons along the $y$ direction with a deflection angle 
\begin{equation}\label{eq:1}
    \Delta\theta_i (\theta',\varphi') = \frac{\lambda_{B} e B_{\mathrm{IGM}} \sin{\theta'}\cos{\varphi'}}{p},
\end{equation}
where $p$ is the momentum of the beam particle and $e$ is the elementary electric charge. $\Delta\theta_i$ is a random variable that depends on the random variables $\theta'$ and $\varphi'$. Since all the possible magnetic field orientation have the same probability, the mean deflection is
\begin{equation}
\begin{split}
    \mu &= \frac{1}{4\pi}\int_0^\pi \sin{\theta'} d\theta' \int_0^{2\pi}d\varphi' \Delta\theta_i(\theta',\varphi') P(\Delta\theta_i(\theta',\varphi'))\\ &= \frac{\lambda_B e B_{\mathrm{IGM}}}{4\pi p}\int_0^\pi d\theta' \int_0^{2\pi}d\varphi' \sin^2{\theta'}\cos{\varphi'} = 0, 
\end{split}
\end{equation}
and the variance is
\begin{equation}
\begin{split}
    \sigma^2 &= \frac{1}{4\pi}\Big(\frac{\lambda_B e B_{\mathrm{IGM}} }{p}\Big)^2 \int_0^\pi d\theta' \int_0^{2\pi}d\varphi' \sin^3{\theta'}\cos^2{\varphi'} \\ &= \frac{1}{3}\Big(\frac{\lambda_B e B_{\mathrm{IGM}} }{p}\Big)^2.
\end{split}
\end{equation}
The total deflection of the beam is computed as
\begin{equation}
    \Delta\theta = \sum_{i=0}^{n} \Delta\theta_{i},
\end{equation}
where $n=\lambda_\text{IC}/\lambda_B$ is the total number of the correlation lengths crossed by the beam during the its energy loss length (substituted here by the inverse Compton scattering length). Since $n$ is very large in our case, $\lambda_\text{IC}>>\lambda_B$, we can use the central limit theorem with $n\xrightarrow[]{}\infty$ to find the distribution function of the total deflection, %. The random variable $Z$
%\begin{equation}\label{eq:lct}
%    Z=\sqrt{n}(\Bar{\Delta\theta_i} - \mu) = \sqrt{n}(\frac{1}{n}\sum_{i=1}^n\Delta\theta_{i}) = %\frac{\Delta\theta}{\sqrt{n}}
%\end{equation}
\begin{equation}\label{eq:th_y3}
    f_{b,\theta_y}(\theta_y,p) = \frac{1}{\sqrt{2\pi\,n}\sigma}
    \exp{-\frac{1}{2}\left(\frac{\theta_y }{\sqrt{n}\sigma}\right)^2}, \text{             } -\pi\leq\theta_y\leq\pi ,
\end{equation}
which is a normal distribution with dispersion $\sqrt{n}\sigma$. 
For the magnetic-field component $B_y = B_{\mathrm{IGM}} \sin{\theta'}\cos{\varphi'}$, we  get the same distribution for $f_{b,\theta_x}(\theta_x,p)$,
\begin{equation}\label{eq:th_x3}
    f_{b,\theta_x}(\theta_x,p) = \frac{1}{\sqrt{2\pi\,n}\sigma}\exp{-\frac{1}{2}\left(\frac{\theta_x}{\sqrt{n}\sigma}\right)^2}, \text{             } -\pi\leq\theta_x\leq\pi. 
\end{equation}
Note that by definition $n=\lambda_\mathrm{IC}/\lambda_\mathrm{B}$. Combining the two distributions in eq.\ref{eq:th_y3} and eq.\ref{eq:th_x3} using the result in appendix \ref{app:transf} we get the full angular distribution of the pair beam
\begin{equation}\label{eq:final}
    f_{b,\theta}(\theta,p) = \frac{1}{\Delta\theta_\text{IGMF}^2\pi} \exp{-\Big(\frac{\theta}{\Delta\theta_\text{IGMF}}\Big)^2}; \text{             } 0\leq\theta\leq\pi, \text{             } 0\leq\varphi\leq2\pi,
\end{equation}
where
\begin{equation}\label{eq:thIMGF}
    \Delta\theta_\text{IGMF}= \frac{eB_\text{IGMF}}{p}\sqrt{\frac{2}{3}\lambda_\text{IC}\lambda_\text{B}}.
\end{equation}

What we have considered here is a fixed IGMF amplitude. Considering a IGMF with different amplitudes leads to the same result in terms of the root-mean-square IGMF with a numerical factor.

\section{Transformation of $f_x(\theta_x)f_y(\theta_y)$ to $f(\theta,\varphi)$}\label{app:transf}

\begin{figure}
    \centering
    \renewcommand{\thefigure}{5}
    \includegraphics[width=5in]{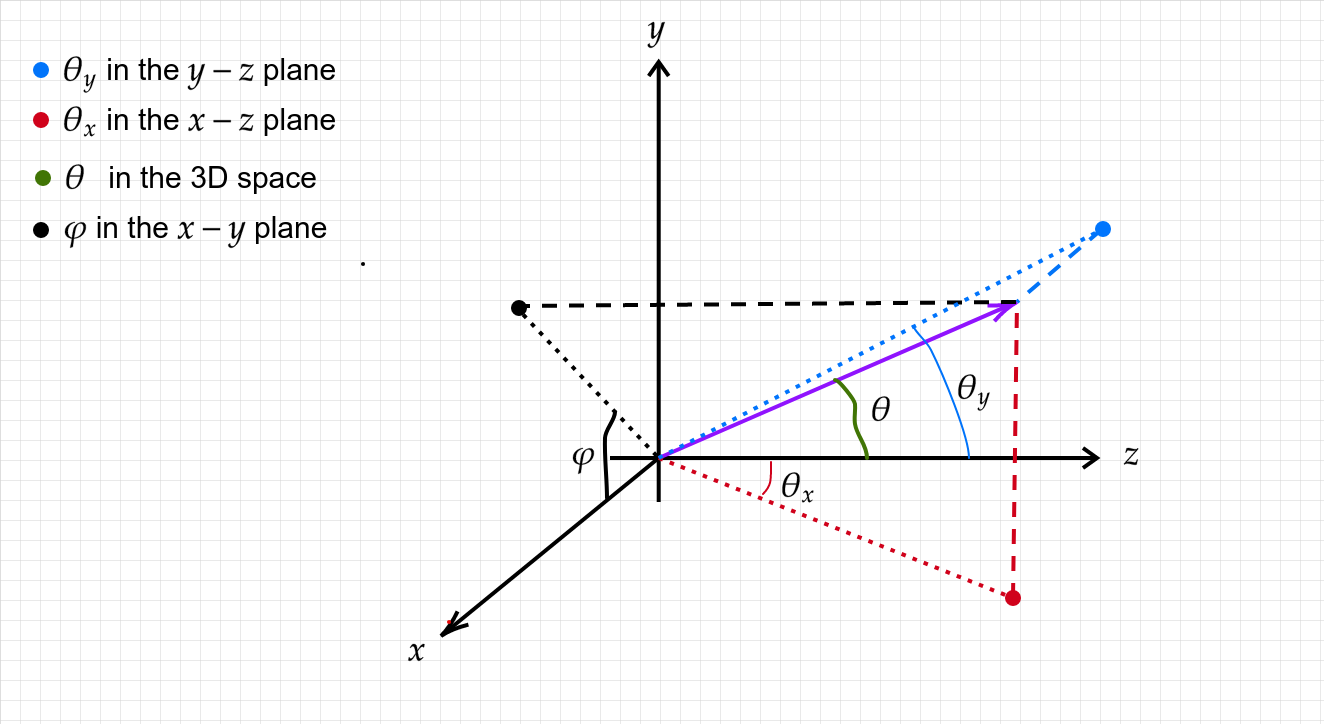}
    \caption{The two systems of coordinate for the angular discretion; the first one uses spherical coordinates, $(\theta,\varphi)$, and the second one is involves $(\theta_x,\theta_y)$, where $\theta_y$ is the angle between the $z$ axis and the projection on the $y-z$ plane and $\theta_x$ is the angle between the $z$ axis and the projection on the $x-z$ plane.}
    \label{fig:coord}
\end{figure}

Combining the distributions $f_x(\theta_x)$ and $f_y(\theta_y)$ to the distribution $f(\theta,\varphi)$ by
\begin{equation}\label{eq:f}
    f(\theta,\varphi) \sin{\theta} d\theta d\varphi = f_x(\theta_x)f_y(\theta_y)d\theta_xd\theta_y,
\end{equation}
where $\theta$ and $\varphi$ are the spherical coordinate and $\theta_x$ and $\theta_y$ are defined in Fig.\ref{fig:coord}. We rewrite eq.\ref{eq:f} using the Jacobian determinant
\begin{equation}\label{eq:J}
    f(\theta,\varphi) = f_x(\theta_x(\theta,\varphi))f_y(\theta_y(\theta,\varphi)) \frac{1}{\sin{\theta}} |\textbf{J}(\theta,\varphi)|.
\end{equation}

The relations between $(\theta_x,\theta_y)$ and $(\theta,\varphi)$ can be found as follows; the Cartesian coordinates of a unit vector with $(\theta,\varphi)$ are $x=\sin{\theta}\cos{\varphi}$, $y=\sin{\theta}\sin{\varphi}$ and $z=\cos{\theta}$, then using the definitions of $\theta_x$ and $\theta_y$ in Fig.\ref{fig:coord} 
\begin{equation}\label{eq:th_xdif}
    \tan{\theta_x} = \frac{x}{z} = \tan{\theta} \cos{\varphi},
\end{equation}
and 
\begin{equation}\label{eq:th_ydif}
    \tan{\theta_y} = \frac{y}{z} = \tan{\theta} \sin{\varphi}.
\end{equation}
The Jacobian determinant is given by
\begin{equation}\label{eq:DJ}
\begin{split}
    |\textbf{J}(\theta,\varphi)| &= \left| \frac{\partial\theta_x}{\partial\theta}\frac{\partial\theta_y}{\partial\varphi} - \frac{\partial\theta_x}{\partial\varphi}\frac{\partial\theta_y}{\partial\theta} \right| \\ &=  \left| \frac{4\tan{\theta}}{4+\sin^2{\theta}\tan^2{\theta}\sin^2{2\varphi}} \right| \approx |\tan{\theta}|,
\end{split}
\end{equation}
for a small $\theta$. Putting this in eq.\ref{eq:J} gives
\begin{equation}\label{eq:re}
    f(\theta,\varphi) = f_x(\theta_x(\theta,\varphi))f_y(\theta_y(\theta,\varphi)) \frac{|\tan{\theta}|}{\sin{\theta}} \approx  f_x(\theta_x(\theta,\varphi))f_y(\theta_y(\theta,\varphi)),
\end{equation}
for a small $\theta$.


\bibliographystyle{aasjournal}
\bibliography{awashra_refs}

%\begin{thebibliography}{99}
%\end{thebibliography}

%%%%%%%%%%%%%%%%%%%%%%%%%%%%%%%%%%%%%%%%%%%%%%%%%%

\end{document}